\documentclass[useAMS,usenatbib]{mn2e}

\usepackage{epsfig}
\usepackage{longtable}
\usepackage{times}
\usepackage{deluxetable}

\def \inte {$INTEGRAL$}
\def \xmm {$XMM$-$Newton$}

\def \sw {$Swift$}
\def \chandra {$Chandra$}

\def \src {IGR\,J17285--2922}
\def \xte {XTE\,J1728--295}

\def \hcm {\hbox {\ifmmode $ atom cm$^{-2}\else atom cm$^{-2}$\fi}}
\def \arcmin {\hbox{$^\prime$}}
\def \arcsec {\hbox{$^{\prime\prime}$}}
\def \chisq {$\chi ^{2}$}

\def \mdot {\dot{M}}

\def \apj {ApJ}
\def \apjl {ApJL}
\def \apjs {ApJS}
\def \aap {A\&A}

\def \mnras {MNRAS}
\def \araa {ARA\&A}
\def \aaps {A\&AS}
\def \nar {New Astronomy Rev.}

\title[\xmm\ and \inte\ observations of \src/\xte\ in outburst]{\xmm\ and \inte\ observations of  the very faint X--ray transient \src/\xte\ during the 2010 outburst}
\author[L. Sidoli, et al.]{L.\ Sidoli,$^{1}$\thanks{E-mail: sidoli@iasf-milano.inaf.it} A.\ Paizis,$^{1}$ S.\ Mereghetti,$^{1}$ D.~G\"otz,$^{2}$ M.\ Del Santo,$^{3}$ \\
$^{1}$INAF, Istituto di Astrofisica Spaziale e Fisica Cosmica,
	Via E.\ Bassini 15,   I-20133 Milano,  Italy \\
$^{2}$AIM (UMR 7158 CEA/DSM-CNRS-Universit\'e Paris Diderot) Irfu/Service
d'Astrophysique, Saclay, F-91191 Gif-sur-Yvette Cedex, France  \\
$^{3}$INAF, Istituto di Astrofisica Spaziale e Fisica Cosmica,
       Via Fosso del Cavaliere 100, I-00133 Roma, Italy
}

\begin{document}

\date{}

\pagerange{\pageref{firstpage}--\pageref{lastpage}} \pubyear{2011}

\maketitle

\label{firstpage}

\begin{abstract}
We report the first broad-band (0.5--150 keV) simultaneous X--ray observations
of the very faint X--ray transient \src/\xte\ performed with \xmm\ and \inte\ satellites
during its last outburst, started on 2010, August 28.
\xmm\ observed the source on 2010 September 9--10, for 22~ks.
\inte\ observations were part of the publicly available
Galactic Bulge program, and overlapped with the times covered by \xmm.
The broad-band spectroscopy resulted in a best-fit with an absorbed power law displaying  a photon index, $\Gamma$, of
1.61$\pm{0.01}$, an absorbing column density, N$_{\rm H}$, of (5.10$\pm{0.05}$)$\times$$10^{21}$~cm$^{-2}$,
and a flux of 2.4$\times10^{-10}$~erg~cm$^{-2}$~s$^{-1}$ (1--100~keV), 
corrected for the absorption.
The data did not require either a spectral cut-off (E$_{c}$$>$50~keV) or an additional soft component. 
The slopes of the \xmm\ and \inte\ separate spectra were
compatible, within the uncertainties.
The timing analysis does not show evidence either for X--ray pulsations or for type I X--ray bursts.
The broad band X--ray spectrum as well as the power density spectrum are  
indicative of a  low hard state in a low mass X--ray binary, although nothing conclusive
can be said about the nature of the compact object (neutron star or black hole).
The results we are reporting here allow us to conclude that \src\ is a low mass
X--ray binary,  located at a distance greater than 4~kpc. 
\end{abstract}

\begin{keywords}
X-rays:  individual (\src, \xte)
\end{keywords}

	\section{Introduction\label{intro}}

Very faint X--ray transients (VFXTs) display outburst peak luminosities
in the range $10^{34}$--$10^{36}$~erg~s$^{-1}$ (2--10 keV),
almost two or three orders of magnitude fainter than the
emission typically shown by most Galactic X--ray transients \citep{Wijnands2006}.
This, together with their apparently small duty cycles, suggests that these black holes or neutron
stars in binary systems undergo a very low average accretion rate \citep{King2006}.

To date, about 30 VFXTs are known and they very likely form a non-homogeneous class of objects,
because their only  common feature is the low  luminosity.
About one third  exhibit
type-I X-ray bursts (\citealt{DelSanto2007, DelSanto2010}) 
and can thus be identified with neutron stars accreting matter
from a low mass companion, but the nature of the remaining sources is unknown \citep{Degenaar2009}.

\src\ is a hard transient discovered in the direction of the Galactic bulge 
with \inte\   in 2004 \citep{Walter2004}.
The source underwent an outburst  lasting at least two weeks
with a peak flux of 1.1$\times10^{-10}$~erg~cm$^{-2}$~s$^{-1}$ (20--150 keV) \citep{Barlow2005}.
For an assumed distance of 8~kpc, this 
corresponds to a luminosity of 8$\times$10$^{35}$~erg~s$^{-1}$, 
which  led to classify \src\ as a VFXT.

More recently, \citet{Markwardt2010} reported on a renewed X--ray activity
started on  August 28th, 2010,
from a transient previously named \xte.
Given the positional coincidence, they
suggested that \xte\ and \src\ are the same source.
\inte\ observations confirmed the renewed activity of \xte\
and its association with \src\ \citep{Turler2010}.

Following this outburst, we triggered a  \xmm\ ToO observation,
with the main aim of an in-depth investigation of the nature of this source.
The  observation  was performed
on 2010 September, 9--10,  about 13 days after the on-set of the  outburst.
We also  analyzed  \inte\ data of the source field obtained,  
as part of the publicly available Galactic 
Bulge program\footnote{http://isdc.unige.ch/Science/BULGE/}\citep{Kuulkers2007}, 
overlapping with the \xmm\ observations.

 	 \section{Observations and Data Reduction\label{dataredu}}

\subsection{\xmm\label{xmmredu}}

The \xmm\ Observatory \citep{Jansen2001} carries three
1500~cm$^2$ X--ray telescopes, each with EPIC imaging spectrometers at the focus. 
Two of the EPIC use MOS CCDs \citep{Turner2001} 
and one uses a pn CCD \citep{Struder2001}. 
RGS arrays \citep{DenHerder2001} are located behind two of the
telescopes. 

\src\ was observed with \xmm\ on 2010 September 9--10.
EPIC pn operated in Large Window mode, while both MOS cameras were in Full Frame mode,
with all the CCDs in Imaging mode.
Both MOS and pn observations used the medium thickness filter.

\xmm\ data were reprocessed using version 10.0 of the Science Analysis Software (SAS). 
Known hot, or flickering, pixels and electronic noise were rejected.
The background (selected with PATTERN=0 and above 10 keV)
did not show evidence of  flaring activity, so no further temporal selection
was applied, resulting in net exposures times of 18.5~ks and 21.6~ks, respectively for  pn and  MOS.
Extraction radii of
40\arcsec\ and 1\arcmin\ were used for the source spectra, respectively
for the pn and MOS cameras.
Background counts were obtained from similar sized region offset from the source position,
resulting in a net count rate  of 10.65$\pm{0.024}$~counts~s$^{-1}$ in the pn spectrum.
Response and ancillary matrix files
were generated using the SAS tasks {\sc rmfgen} and {\sc arfgen}.
Using the SAS task {\sc epatplot} we found that only the MOS spectra were affected by pile-up.
Thus, we excluded the inner 6\arcsec\ (radius) of the PSF from the MOS1 and MOS2 spectra
adopting only PATTERN=0, while for EPIC pn spectrum we selected PATTERN from 0 to 4.

To ensure applicability of the \chisq\ statistics, the
net spectra were rebinned such that at least 20 counts per
bin were present and such that the energy resolution was not
over-sampled by more than a factor 3. 
All spectral uncertainties and upper-limits given below are at the 90\% confidence level for
one parameter of interest.
We performed the data analysis using
HEASoft 6.10 and XSPEC v.12. 

The RGS was operated in spectroscopy mode  and
resulted in a net exposure of 21.6~ks.
RGS source and background events were calibrated by applying
the latest calibration parameters.

\subsection{\inte\label{interedu}}

We analyzed the data of the Galactic Bulge monitoring
program in
the time-frame close to the source outburst as detected by \textit{RXTE} on
August 28th, 2010.
We used the imager IBIS/ISGRI (\citealt{Ubertini2003}, \citealt{Lebrun2003})
on-board \textit{INTEGRAL} \citep{Winkler2003},
and analyzed a total of 68 pointings, each
with an exposure time of about 1800\,s (nominal), spanning from August 26, 2010
to October 6, 2010.

\begin{figure}
\begin{center}
\centerline{\includegraphics[width=8cm,angle=0]{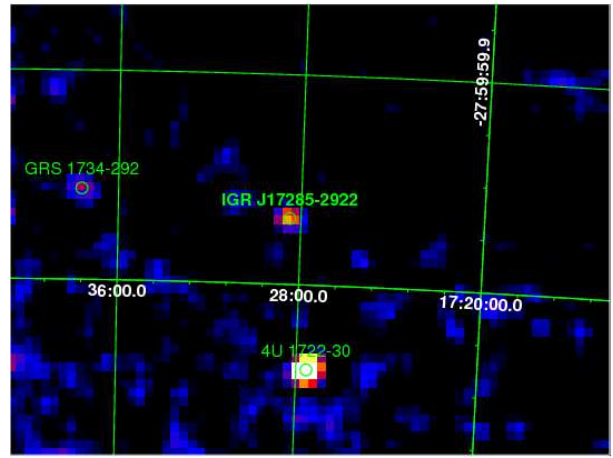}}
\caption{ISGRI significance map obtained in the 30.2--50.3\,keV band.
IGR~J17285--2922 is detected with a  significance of about 7.5}
\label{isgri}
\end{center}
\end{figure}

Version 9.0 of the Off-line Scientific Analysis (OSA) software was used to
analyse the data. For each pointing we extracted images in the \mbox{17.8--25,
25--30.2, 30.2--50.3,50.3--80,80--150.4\,keV} energy bands (the boundaries 
have been chosen in order to cope with the response matrix).
The images were used to build
one-pointing based light-curves as well as a final mosaic. The source has never
been detected at a single pointing level, or at a one mosaic per revolution
level, but it is clearly detected in the total mosaic (see Fig~\ref{isgri}).
We extracted an average spectrum from the total mosaic using the {\sc mosaic\_spec}
tool available within the OSA~9.0 software package.

  	\section{Analysis and Results\label{result}}

\subsection{Light curves and Timing Analysis}

\begin{figure}
\begin{center}
\centerline{\includegraphics[width=7cm,angle=-90]{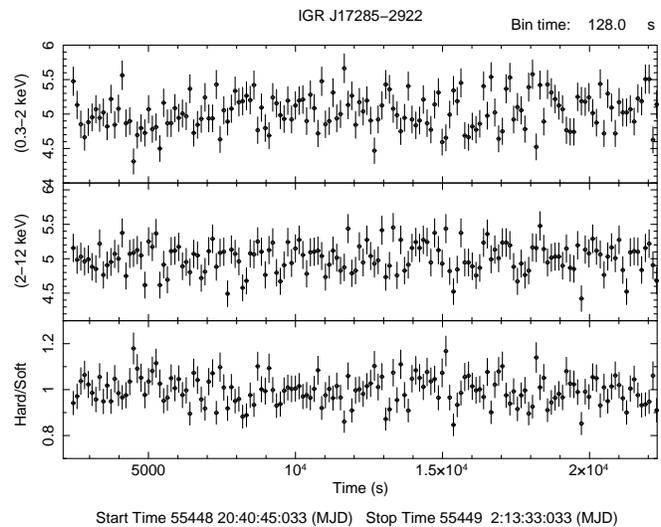}}
\caption{\src\ light curve with
EPIC pn in two energy ranges (below and above 2 keV) and
their hardness ratio. Bin time is 128 s.}
\label{pnlc}
\end{center}
\end{figure}

\begin{figure}
\begin{center}
\centerline{\includegraphics[width=6.0cm,angle=-90]{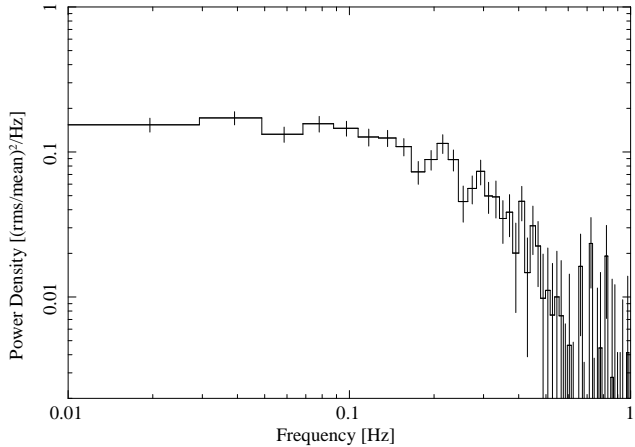}}
\caption{Power spectrum of EPIC pn source events in the energy range 0.3--12 keV.}
\label{timing}
\end{center}
\end{figure}

The light curves of \src\ observed with the EPIC pn in
the soft (0.3--2 keV) and hard (2--12 keV) energy ranges 
are shown in Fig.~\ref{pnlc}. 
A similar behaviour is displayed by source emission observed with both MOS1 and MOS2.
The average flux does not vary during the observation, but some rapid
variability is present. 
This is clearly  visible in the power spectrum shown in
Fig.~\ref{timing}, which has  been obtained by averaging the power
spectra (0.3--12 keV) of 391 time intervals of 51.2~s each,
binned at 0.1~s. The fractional rms variability, integrated over the
0.01--1 Hz range, is about 20\%.

The hardness ratio between the soft and hard energy ranges (bottom panel of
Fig.~\ref{pnlc}) is consistent with a constant value.
A fit with a constant gives a  value of 
$0.989\pm0.004$ ($\chi^2_{\nu}=1.27$ for 156 d.o.f.).
Therefore in the following we
perform a spectral analysis integrating over the whole duration
of the observation.

The low statistics hampers a meaningful temporal analysis at hard energies, 
since the source is not detected in single pointings, but only 
in the total mosaic of the summed IBIS/ISGRI observations (see Section~\ref{interedu}).

\subsection{Spectroscopy}\label{section:spec}

Fitting the EPIC spectra (pn+MOS1+MOS2) with an absorbed power law
resulted in structured residuals near 2.2 keV and below 1 keV,
which can be ascribed to residual uncertainties in the calibration.
The largest departure of the data with respect to the  model
is due to narrow negative residuals near 2.2 keV present in the EPIC pn, 
likely due to an incorrect instrumental modeling of the gold mirror edges,
as already noticed, for example, in the \xmm\ spectrum of  GRO~J1655-40 \citep{DiazTrigo2007}.
The other discrepancies present around and below 1 keV have often been observed
in other  X--ray binaries, especially in case of hard X--ray emission
(e.g. \citealt{Boirin2005}, \citealt{Sidoli2005}, \citealt{Sidoli2008}).
Some authors usually exclude the softest part of EPIC data, other
include Gaussian lines to account for these residuals (e.g \citealt{DiazTrigo2007}).
Here we decided to not exclude particular energy ranges, but instead to include a 2\% systematic 
error in EPIC data, both when fitting the EPIC spectra alone (pn+MOS1+MOS2), and when
fitting them together with ISGRI higher energy spectrum.
Other authors adopted similar or even higher systematic errors \citep{CadolleBel2004}
to account for these residual discrepancies.
Note however that, if we perform the spectroscopy only considering the 
higher energy range  2.5-10 keV (EPIC data), the 
resulting spectral parameters are always consistent with the results we are reporting in the following paragraphs.

The spectroscopy of the EPIC data alone (0.5--10 keV) with an
absorbed power law ($\chi^2_{\nu}=1.094$ for 640 d.o.f) resulted in the following parameters:
$N_{\rm H}$=(5.10$\pm{0.05})$$\times$$10^{21}$ cm$^{-2}$,
photon index, $\Gamma$, of 1.61$\pm{0.01}$.
Adopting alternative simple models resulted in much worse fits:
a multicolour disk blackbody ({\sc diskbb} in {\sc xspec}) or a simple blackbody
gave $\chi^2_{\nu}$$>$5.
Additional soft components to the power law model, as well as a high energy exponential 
cut-off, were not required by the data.
Other complex fits, e.g. a multicolour disk blackbody ({\sc diskbb} in {\sc xspec})
plus a blackbody, although formally acceptable ($\chi^2_{\nu}$=1.100 for 638 d.o.f.),
underestimated the flux seen  higher energies with \inte, thus requiring  an additional
hard power law component.

The RGS spectra (0.5--2 keV) resulted in  net source count rates of 0.150$\pm{0.003}$~counts~s$^{-1}$
and 0.192$\pm{0.003}$~counts~s$^{-1}$ respectively in RGS1 and RGS2.
RGS spectra did not show evidence for narrow lines.
An absorbed power law was a good fit to the data ($\chi^2_{\nu}=0.918$ for 3038 d.o.f.)
resulting in a column density in the range [0.57--0.77]$\times$$10^{22}$ cm$^{-2}$
and a photon index, $\Gamma$, between 1.48 and 2.10 (90\% uncertainty).
This is consistent with the EPIC results, so we will not discuss the RGS data further.

The \inte/ISGRI spectrum (17.8--150.4 keV)  displayed a slope
consistent with the one seen with \xmm\ below 10 keV: a fit with
a power law resulted in a photon index of 1.7$\pm{0.3}$ ($\chi^2_{\nu}=0.742$ for 3 d.o.f.).

\begin{figure}
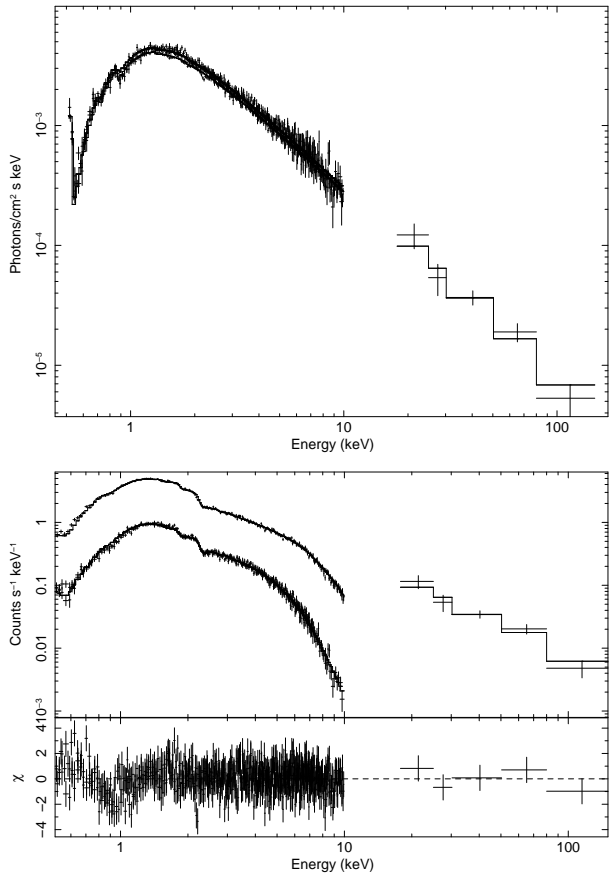

\begin{center}
\begin{tabular}{cccc}
\includegraphics[height=8.0cm,angle=-90]{ufs_pow.ps} \\
\includegraphics[height=8.0cm,angle=-90]{ldadelchi_pow.ps}
\end{tabular}
\end{center}
\caption{0.5--150 keV broad band spectrum of \src\ (pn+MOS1+MOS2 together with ISGRI data), 
fitted with an absorbed power law model.
{\em Upper panel} shows the photon spectrum, while the {\em lower panel} displays
the counts spectrum together with the residuals in units of standard deviations.}
\label{spec}
\end{figure}

\begin{figure}
\begin{center}
\centerline{\includegraphics[width=6.5cm,angle=-90]{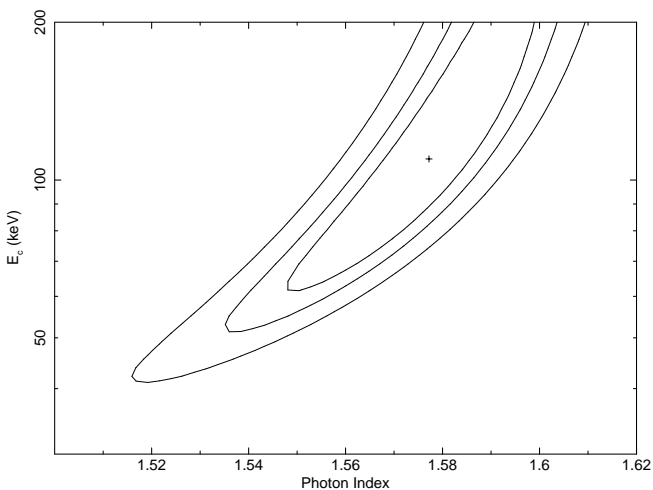}}
\caption{Confidence contour levels (68\%, 90\%, 99\%) for the two parameters
high energy cut-off, E$_{\rm c}$,  and photon index, when fitting the broad band spectrum
with an absorbed cut-off power law ({\sc cutoffpl} model in {\sc xspec}).}
\label{cont2}
\end{center}
\end{figure}

We next analysed the broad band 0.5--150 keV emission with a  joint fit of  \xmm/EPIC (pn+MOS~1+MOS~2) and \inte/ISGRI. 
We included constant factors in
the spectral fitting to allow for normalization
uncertainties between the instruments.
An absorbed power law model resulted in a good fit 
($\chi^{2}_{\nu}$/d.o.f.=1.091/644), 
as shown in Fig.~\ref{spec}.
The best-fit spectral parameters are equal to those obtained for the EPIC spectrum alone
and the ISGRI/EPIC pn constant factor was 1.17$\pm{0.18}$.
The fluxes corrected for the absorption are the following:
F=6.8$\times10^{-11}$~erg~cm$^{-2}$~s$^{-1}$  and
F=2.4$\times10^{-10}$~erg~cm$^{-2}$~s$^{-1}$,
respectively in the 1--10 keV and 1--100 keV energy ranges 
(assuming the EPIC pn response matrix extrapolated to higher energies).
Fitting the EPIC and ISGRI data  with a power law with a high-energy exponential cut-off
({\sc  cutoffpl} model in {\sc xspec})  allows us to put
a lower limit to the cut-off energy, E$_{\rm c}$, of $\sim$50~keV (Fig.~\ref{cont2}, 90\% confidence level).

We also tried a double component model, adding a soft component to the  power law continuum: 
using a blackbody together with a power law,
we obtained a blackbody temperature of 0.61$^{+0.86} _{-0.07}$~keV and 
a small radius  R$_{bb}$ = 0.80 $^{+0.46} _{-0.38}$~km (at a distance of 8~kpc; $\chi^{2}_{\nu}$/d.o.f.=1.082/642).
An F-test 
resulted in a probability of $2.631\times10^{-2}$.
Similar risults were obtained assuming a {\sc diskbb} model for the additional
component ($\chi^{2}_{\nu}$/d.o.f.=1.080/642; F-test probability of $1.187\times10^{-2}$),
resulting in an inner disc temperature of 1.2 $^{+2.0} _{-0.3}$~keV 
and in an innermost disc radius, R$_{in}$$\times (cos~i)^{0.5}$, of 0.20$^{+0.31} _{-0.09}$~km at 8 kpc 
($i$ is the disc inclination).
Therefore we conclude that, even if we cannot rule out the presence of 
a weak additional soft component, there is 
no statistical evidence for its presence in the current data.

We next tried to describe  the broad-band spectrum 
with physical models which involve a Comptonizing plasma, like
{\sc compTT} and {\sc bmc} in {\sc xspec}.
A fit with the {\sc compTT} model  \citep{Titarchuk1994} returns 
a scenario with cold seed photons (0.05--0.09\,keV) 
upscattered by a corona (disc geometry) 
with optical depth $\tau$=1.5$^{+0.5}_{-1.4}$ and electron temperature kT$_{e}$$>$20\,keV 
($\chi^{2}_{\nu}$/d.o.f.=1.080/642).

We also fit the data with the  {\sc bmc} model \citep{TMK1996}.
This model is the sum of a blackbody ({\sc bb}) plus its Comptonization, the latter 
obtained as a consistent convolution of the {\sc bb} itself with Green's 
function of the Compton corona. The free parameters of the {\sc bmc} model 
(apart from the normalization) are the {\sc bb} colour temperature, kT$_{\rm BB}$, 
the spectral energy index, $\alpha$, and the logarithm of the 
illuminating factor A, log~A.  The log~A parameter is an indication of the 
fraction of the up-scattered {\sc bb} photons with respect to the {\sc bb} seed 
photons directly visible. 
In the extreme cases, the seed photons can be 
completely embedded in the Comptonizing cloud (none directly visible, 
A$>>$ 1, e.g. log~A=8) or there is no coverage by the Compton cloud 
(A$<<$1, e.g. log~A=$-8$), and we directly observe the seed photon spectrum 
(equivalent to a simple {\sc bb}, with no Comptonization). 
In our case we obtain ($\chi^{2}_{\nu}$/d.o.f.=1.027/642)
a kT$_{\rm BB}$=0.07$^{+0.03 }_{-0.01}$\,keV, seed photon 
population up-scattered  with $\alpha$=0.64$\pm{0.02}$ and log(A)=$-$0.24$^{+0.02 }_{-0.1}$.
We note that the {\sc bmc} model has no cut-off in it (i.e. we are in the 
power-law shape case) and the well constrained $\alpha$ parameter 
($\Gamma$=$\alpha$+1) indicates the overall Comptonization efficiency 
related to an observable quantity in the photon spectrum of the data (the 
slope $\alpha$, unlike kT$_{e}$ and $\tau$ that are not directly 
observable in the spectrum).  
The lower the  $\alpha$ value, the higher 
the efficiency, that is, the higher the energy transfer from the hot 
electrons to the soft seed photons.

	\section{Discussion and Conclusions\label{discussion}}

Our \xmm\ ToO observation of \src/\xte\ triggered by its recent outburst, coupled
with simultaneous INTEGRAL data, allowed us to derive the first broad band spectrum
(0.5--150 keV) of this VFXT.
During this outburst, the second shown by this source in almost seven years, follow-up observations
were carried out with different satellites and ground based telescopes. 
The source position was first refined thanks to a \sw/XRT pointing \citep{Yang2010:atel2824}.
This  ruled out all the  six  sources detected  with \chandra\ in the \inte\
error circle  when \src\ was in quiescence \citep{Tomsick2008} as possible soft X--ray counterparts. 
A sub--arcsecond position was  later determined with \chandra\ \citep{Chakrabarty2010},
leading to the identification of a likely optical counterpart 
(\citealt{Russell2010:atel2827}, \citealt{Torres2010},
\citealt{Russell2010:atel2997},
\citealt{Kong2010}).
This star, at coordinates
R.A. (J2000) = 17$^h$~28$^m$38.86$^s$,
Dec (J2000) = $-29^\circ$~21$'$~44.0$''$ \citep{Torres2010},
appeared bluer and more variable than other candidates inside the \chandra\ error region.
It was not detected in archival optical images taken three months before the last outburst,
with an upper limit of R-magnitude$>$21  \citep{Kong2010}.

The faintness of the optical counterpart in quiescence
allows us to better constrain  the source nature and its distance.
In the following, we use a visual extinction A$_V$=2.4 mag (which implies A$_R$=1.8 mag), derived 
from the absorbing column density resulting 
from the  X--ray spectroscopy \citep{Guver2009}.

A HMXB can be excluded, because, even if placed at the
Galactic boundaries, it would have a brighter R magnitude.
For example, to have  R$>$21~mag, a  B0V star should lie
at a distance larger than 450~kpc, and a B0.5 supergiant star
at more than 1.6~Mpc.
The source is  more likely a LMXB, being fully
compatible with the observed constraint: for example, a K5V companion
star (assuming M$_V$=$+7.3$, V$-$R=$+0.99$; \citealt{Johnson1966}),
placed at  8~kpc, would show a magnitude R$\sim$22--23.
On the other hand, the measured upper limit R$>$21~mag would
imply a LMXB distance larger than $\sim$4~kpc.
Thus, we conclude that \src\ is a LMXB located at a distance
greater than 4~kpc.

\begin{figure}
\begin{center}
\centerline{\includegraphics[width=7.0cm,angle=-90]{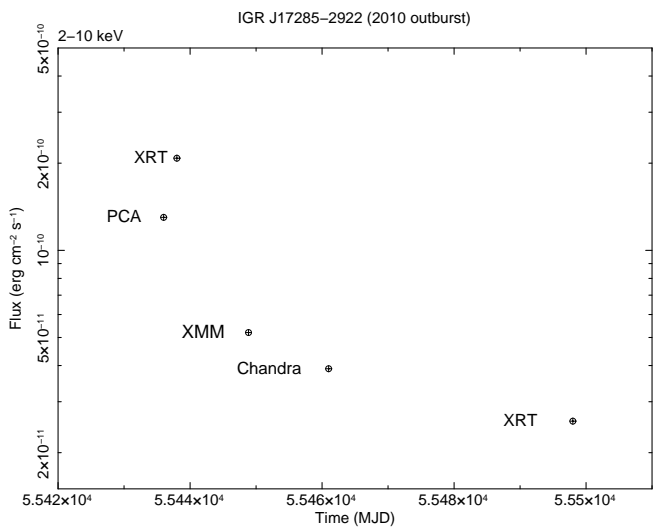}}
\caption{Evolution of the 2010 outburst. Absorbed fluxes in the energy range 2--10 keV
have been calculated from the best-fit parameters reported in Table~\ref{tab:history}, except for the
\xmm\ observation, where we show the observed 2--10 keV flux obtained from our EPIC spectroscopy.}
\label{fig:history}
\end{center}
\end{figure}

\begin{table*}
\begin{center}
\caption{Summary of the published observations of \src\ performed during the 2010 outburst.}
\label{tab:history}
\begin{tabular}{lccccc}
\hline
Time         &    Energy band       &   Unabsorbed Flux                                &  Power law      &     Column density  &           Refs.    \\
(YYYY-MM-DD) &        (keV)         &   $10^{-10}$~erg~cm$^{-2}$~s$^{-1}$   &          $\Gamma$           &    $10^{21}$~cm$^{-2}$~               \\
\hline
2010-08-28   &     2--10            &   1.3                                 &            $-$           &   $-$  &  \citet{Markwardt2010}   \\
2010-08-30   &     0.3--10          &   6.1                                 &     2.23$\pm{0.26}$      &   5.4$\pm{1.2}$   & \citet{Yang2010:atel2824}  \\
2010-08-26/30 &     20--80          &   0.98                                &     2.1$\pm{0.5}^{a}$    &   $-$    & \citet{Turler2010}   \\
2010-09-09/10 &     2--10  (0.3--10) &   0.54  (0.86)                       &     1.61$\pm{0.01}$      &   5.10$\pm{0.05}$   & {\em This work}    \\
2010-09-22   &     0.1--10           &   1.6                                &         2.23 fixed       &   5.4 fixed  & \citet{Chakrabarty2010}   \\
2010-10-29   &     0.3--10           &   0.46                               &     1.7$\pm{0.2}$        &    6.2 $^{+1.4} _{-1.2}$ & \citet{Yang2010:atel2991}     \\
\hline
$^{a}$ 1~$\sigma$ error. \\
\end{tabular}
\end{center}
\end{table*}

During its first outburst in 2003, \src\ was caught at first
in a soft state with the $RXTE$ satellite,
consistent with a steep power law with a photon index of 3.6--3.8 \citep{Markwardt2010}.
Then, during \inte\ observations performed about one month later,
the 20--150 keV spectrum seemed to be harder, with $\Gamma$=2.1$\pm{0.17}$ \citep{Barlow2005}.

During the evolution of the second outburst in 2010 
(see Table~\ref{tab:history} and Fig~\ref{fig:history}), 
the source spectrum was apparently harder when fainter, with a continuum
always dominated by a power law with a slope
within the  canonical range for the low-hard state in LMXBs 
($\Gamma$$\sim$1.5--1.7 for BH binaries, hereafter BHB, \citet{Belloni2010}).
This behavior (low-hard state during the entire outburst) is consistent with a BH nature,
although the canonical  evolution
of the outburst in a BH transient (BHT) starts with a low-hard state
and then undergoes a transition to a high-soft
state, where the thermal emission from the accretion disc dominates
the X--ray spectrum, following a  q-shaped behavior
in the hardness-intensity diagram (see the recent review \citealt{Belloni2010} and references therein).
However, not all BHTs  go through low-hard to high-soft states during 
their outbursts, but a few of them remain in the low-hard state until they return to quiescence 
(see, e.g., \citealt{Brocksopp2004}, \citealt{Capitanio2009}).

The \src\ broad-band X-ray spectrum we have reported here, if fitted with physical
models (see Section~\ref{section:spec}), draws a scenario compatible with the typical
low hard state of a LMXB (cold and distant disc Comptonized by a hot corona),
but little can be said from the spectral point of view on the nature of
the compact object, since a plasma temperature of kT$_{e}$$>$20\,keV 
(and $\Gamma$=$\alpha$+1=1.6) 
has been observed in both BH and neutron star LMXBs 
(e.g. \citealt{Paizis2006}, \citealt{Bouchet2009}, \citealt{Cocchi2010}).

Observations suggest that quiescent BHTs  as a class are fainter than  transients
containing a neutron star with similar orbital periods  (\citealt{Garcia2001}, \citealt{Narayan2008}).
Tomsick et al. (2008)  found six faint sources within the \inte\ error circle during
a \chandra\ observation when the source was in quiescence,
thus leading to an unclear counterpart. Taking the brightest
of these faint \chandra\ sources, they calculated a conservative upper limit
to the X--ray emission in quiescence of 5.5--6.4$\times$10$^{-14}$~erg~cm$^{-2}$~s$^{-1}$
(0.3--10 keV unabsorbed flux, assuming power law photon indices between $\Gamma$=1 and 2).
A refined error circle allowed to exclude  \citep{Yang2010:atel2824},
as possible soft X--ray counterparts, all these  six   \chandra\ sources.
Thus we can re-calculate these upper limits to the quiescent emission,
rescaling these fluxes at least to the faintest \chandra\ source in the error circle,
resulting in a new upper limit of 
2.4--2.8 $\times$10$^{-14}$~erg~cm$^{-2}$~s$^{-1}$ (0.3--10 keV, unabsorbed flux,
assuming $\Gamma$=1--2).
This translates into an X--ray luminosity
in quiescence L$_{\rm quiesc}$$<$(1.7--2.0)$\times$(d$_{\rm 8 kpc}$$^2$)$\times$10$^{32}$~erg~s$^{-1}$,
where d$_{\rm 8 kpc}$  is the source distance in units of 8~kpc.
If \src\ is located closer than the Galactic Centre, this conservative upper limit to the quiescence
becomes low and possibly indicative of a BHB.

Neither X--ray pulsations nor type I X--ray bursts have been observed.
The power density spectrum (PDS)  measured with \xmm\ resembles the
typical shape and normalization of aperiodic variability
in low-hard states of LMXBs  (\citealt{McClintock2006}, \citealt{Belloni2010}), but nothing really conclusive
can be said about the nature of the compact object.

This observation, although leading to the first broad-band spectroscopy up
to 150~keV, demonstrated that it is very difficult to discriminate a black hole
from a neutron star in a VFXT. From the spectral point of view, it does
not exist, to date, a firm spectral signature which allows to distinguish a 
black hole from a neutron star, especially if the X--ray transient remains 
in a low/hard state along the entire outburst. 
The same can be said about the power density spectrum
in the frequencies range of our data: a possible way to distinguish 
a black hole from a neutron star was proposed by \citet{Sunyaev2000},
but it involves power density spectra at higher frequencies, above 500~Hz.
Moreover, in  VFXTs ($\mdot < 2 \times10^{-10}$~M$_\odot$~yr$^{-1}$) hosting accreting neutron stars,
the type I X--ray bursts seem to be rare (\citealt{Cornelisse2004}, \citealt{Wijnands2008}), although
a proper comparison with the different possible explanations (e.g. \citealt{Peng2007})
needs more observational data, especially on the bursts recurrence time at different accretion
regimes.

\section*{Acknowledgments}

This work is based on data from observations with \xmm, and \inte.
\xmm\ is an ESA science mission with instruments and
contributions directly funded by ESA Member States and the USA (NASA).
\inte\ is an ESA project with instruments and
science data centre funded by ESA member states
(especially the PI countries: Denmark, France, Germany, Italy, Switzerland, Spain),
Czech Republic and Poland, and with the participation
of Russia and the USA.
The \inte\ data used in this paper are taken from the \inte\ Galactic bulge
monitoring program (PI E. Kuulkers) that are publically available and hence offer
the unique opportunity to study broad-band spectra of active sources in the
Galactic bulge, in a fruitful synergy of operating high energy missions.
We thank the \xmm\ duty scientists and science planners for making these observations possible,
in particular Rosario~Gonzalez-Riestra (\xmm\ Science Operations Centre 
User Support Group).
This work was supported in Italy by contracts ASI/INAF I/033/10/0 and I/009/10/0 
and by the grant from PRIN-INAF 2009, 
``The transient X-ray sky: new classes of X-ray binaries containing neutron stars''
(PI: L. Sidoli).

\bibliographystyle{mn2e}

\bsp

\label{lastpage}

\end{document}